\documentclass[nofootinbib,floats,aps,superscriptaddress,preprint]{revtex4}   
\newcommand{\ba}{\begin{array}}   
\newcommand{\ea}{\end{array}}   
\newcommand{\bd}{\begin{displaymath}}   
\newcommand{\ed}{\end{displaymath}}   
\newcommand{\be}{\begin{equation}}   
\newcommand{\ee}{\end{equation}}   
\newcommand{\beq}{\begin{equation}}   
\newcommand{\eeq}{\end{equation}}   
\newcommand{\bea}{\begin{eqnarray}}   
\newcommand{\eea}{\end{eqnarray}}   
   
\def\bra{\langle}   
\def\ket{\rangle}   
   
\def\q2 {q^2}

\def\r {\rightarrow}

 \def\N0{\widetilde \chi^0}

\def\lsim{\:\raisebox{-0.5ex}{$\stackrel{\textstyle<}{\sim}$}\:}

\begin{document}   
   
\preprint{\vbox{\hbox{hep-ph/0204216}\hbox{April, 2002}}}   
   
\vspace*{1cm}   
   
\title{Effects of initial axion production and photon--axion   
oscillation on type Ia supernova dimming}  
  
\author{Yuval Grossman\footnote{yuvalg@physics.technion.ac.il}}   
\affiliation{Department of Physics, Technion--Israel Institute of  
    Technology\\   
       Technion City, 32000 Haifa, Israel\vspace{6pt}}

\author{Sourov Roy\footnote{roy@physics.technion.ac.il}}   
\affiliation{Department of Physics, Technion--Israel Institute of  
    Technology\\   
       Technion City, 32000 Haifa, Israel\vspace{6pt}}   
   
\author{Jure Zupan\footnote{jure.zupan@ijs.si}}   
\affiliation{Josef Stefan Institute\\    
Jamova 39, P.O.Box 3000, SI-1001 Ljubljana, Slovenia   
	\\[20pt] $\phantom{}$ }

\begin{abstract} \vspace*{10pt}   
   
Recently, Cs\'aki, Kaloper and Terning (CKT) suggested that  
axion-photon oscillation in the intergalactic medium can explain the  
observed dimming of distant type Ia supernovae. This mechanism works  
only if the initial axion flux is much smaller than the initial photon  
flux. We study several mechanisms for such initial axion  
production. The direct axion production in the supernovae, the  
photon-axion oscillation in the magnetic field of supernovae and in  
the magnetic field of their host galaxies are addressed. We find it  
likely that the initial axion flux is very small and therefore does not 
pose a problem to the CKT mechanism.  
   
\end{abstract}   
   
\maketitle

\section{Introduction}   
 
Type Ia Supernovae (SNe Ia) are very interesting objects in
observational cosmology \cite{bruno}.  Recent observations of distant
($z \sim 1$) SNe Ia have raised this interest enormously because the
data indicate that they are fainter than would be expected for a
decelerating universe \cite{SCP,HZT}. The dimming seems to be
independent of the wave length of the emitted light. The standard
interpretation of this result is that our universe is accelerating at
present as a result of a non-vanishing cosmological constant. Several
alternatives to this idea have been suggested.  One idea suggests that
the light from distant SNe Ia can be partially absorbed by the dust
which may be present on the light path \cite{dust}.  Ideas involving
extra dimensions have also been suggested \cite{extradim}.  The effect
of cosmic evolution on distant SNe Ia was studied in \cite{evolve}.
   
Recently, Cs\'aki, Kaloper and Terning (CKT) \cite{ckt1} proposed  
another explanation of the dimming due to oscillation of photons into  
very light axions in the background magnetic field of the  
intergalactic plasma. The statistical significance of this result and  
possible generalizations of the mechanism are discussed in  
Ref. \cite{erlich}.

The intergalactic plasma produces an energy dependent photon--axion
oscillation and thus a frequency dependent dimming 
\cite{deffayet,ckt2,MBG}. This is in contrast to the achromatic dimming 
caused by a non-vanishing cosmological constant. While in the future the 
energy dependence may be the best tool to discriminate between these two 
mechanicsms, the present data are not sensitive enough \cite{ckt2,MBG}. 
Based on this fact, it is important to further investigate the CKT mechanism. 

In all these works \cite{ckt1,erlich,deffayet,ckt2,MBG} it was  
implicitly assumed that only photons are emitted from the supernova  
(SN) while axions are produced only via oscillation in the  
intergalactic medium. If this assumption failed the dimming would be  
less significant.  As an example consider the case of  
polarized photons, where the initial axion flux is equal to the  
initial photon flux (with polarization state parallel to the external  
magnetic field). Then, since the photon to axion and axion to photon  
oscillation probabilities are equal ($P_{\gamma \r a}$ = $P_{a \r  
\gamma}$) the oscillation will cause no effect at all. More generally,  
in the case of unpolarized photons and/or randomly distributed  
background magnetic field, the photon intensity on Earth is given  
by  
\be   
I_\gamma(y)=I_\gamma(0)-P_{\gamma \r a}(y) \left[I_\gamma(0)-    
2 I_a(0)\right]   
\ee   
where $y$ is the distance traveled by the photons, $I_\gamma(0)$  
$[I_a(0)]$ is the initial photon [axion] flux, $I_\gamma(y)$ is the  
final photon flux and $P_{\gamma \r a}(y)$ is the oscillation  
probability.  Clearly, if the initial axion flux is significant it  
must be taken into account.  
   
In this paper we estimate the initial axion flux for  the relevant   
energy range. We consider axion production in the SN, axion conversion in   
the SN magnetic field and axion conversion in the magnetic field of   
the host galaxy. We find that it is unlikely that the initial axion   
flux is significant. We thus conclude that neglecting the initial   
axion flux does not pose a problem to the axion--photon mixing   
explanation of the dimming.   
   
\section{Photon-axion mixing formalism}   
 
In a background magnetic field axions and photons oscillate into each other   
\cite{sikivie,raffelt} due to the axion--photon coupling   
\be   
{\cal L}_{int} = {a \over M} \vec E \cdot \vec B,   
\ee   
where $a$ is the axion field, $\vec E$ is the electric field of the
photon, $\vec B$ is the background magnetic field and the mass scale
$M$ characterizes the strength of the axion-photon interactions.  In
Ref. \cite{ckt1} it was found that $M=4 \times 10^{11}\,$GeV gives the
best fit to the data and we will use this value in the analysis.  In
order to have mixing a transverse external magnetic field is
needed. Moreover, only the plane wave photon polarization state
parallel to the external field can mix with the axion.
   
Photon--axion oscillation is induced by the following mass-squared
matrix
\be   
{\cal M}^2 = \pmatrix{\omega^2_p & i {\cal E} \mu \cr   
-i {\cal E} \mu & m^2 \cr} \,.\label{matrix}   
\ee   
where $\cal E$ is the photon energy,   
\beq   
\mu = {B\over M}\,,   
\eeq   
$B \sim |{\bf B}|$ is the projection of the magnetic field on the  
direction of the photon polarization and $m$ is the axion mass. In  
Ref.  \cite{ckt1}, the authors found that $m\sim 10^{-16}\, {\rm eV}$  
is needed. For the purposes of this paper, as long as 
$m^2 \ll \max({\cal E} \mu,\omega^2_p)$,  we do not need the exact  
value of the axion mass ($m\sim 10^{-16}\, {\rm eV}$ is small enough).
The plasma frequency is  
given by  
\be   
\omega^2_p = {{4 \pi \alpha n_e} \over {m_e}} \,,  
\label{plasma}   
\ee   
where $m_e$ is the electron mass and $n_e$ is the number density    
of electrons in the plasma. We omitted the Euler-Heisenberg term    
\cite{heisenberg}   
$\omega^2_{EH} = (7 \alpha/ 90 \pi)(B / B_{cr})^2{\cal E}^2$   
(where $B_{cr} = 4.41 \times 10^{13}~{\rm G}$ is the critical field  
strength \cite{raffelt}), which contribute to ${\cal M}^2_{11}$, since  
the magnetic fields we are considering are much smaller than $B_{cr}$  
and the effect of this term is negligible.  
   
For a constant magnetic field the oscillation probability of photons    
into axions is given by   
\be \label{werww}   
P_1 = {\mu^2 \over k^2} \sin^2\left({k x \over 2}\right) =    
{(\mu x)^2\over 4}\left(\sin({k x/2}) \over {k x/2}\right)^2 \,,  
\label{probab}   
\ee   
where $x$ is the distance traveled by the photon,   
$k=2\pi/ L_{Osc}$ and the oscillation length is given by   
\be   
L_{Osc} = \frac{4\pi {\cal E}}{\sqrt{{(\omega^2_p - m^2)}^2    
+ 4 \mu^2 {\cal E}^2}} \,.   
\ee   
Note that the oscillation probability is bounded by    
\beq \label{onepro}   
P_1 \le {(\mu x)^2\over 4}\,.   
\eeq   
Of particular interest to us is the case where the constant magnetic 
field domain size ($L_{dom}$) is small, namely when $kL_{dom}\ll 1$. 
Expanding in this small parameter one finds that the bound of 
Eq. (\ref{onepro}) (with $x=L_{dom}$) is saturated, namely $P_1=(\mu 
L_{dom})^2/ 4$. In particular, in this case the conversion is 
independent of the axion mass, the photon energy and the plasma 
frequency. This is the reason that the CKT mechanism does not 
produce energy dependent dimming as long as the plasma frequency is 
small enough such that the above approximation is valid \cite{ckt2}. 
   
Next we consider the case where the magnetic field is not constant.   
Within a set of simplifying assumptions (see the Appendix for the  
derivation) the total photon-axion conversion over many domains   
is given by   
\be \label{totpro}   
P_{\gamma \leftrightarrow a}(y) = {1 \over 3}  
\left(1-e^{-y/L_{decay}}\right)\,,  
\ee   
such that $y=N L_{dom}$ is the total length traveled by the photons in   
$N$ domains and   
\beq   
L_{decay}={2 L_{dom} \over 3 P_1},    
\eeq   
where $P_1$ is given in (\ref{werww}). Using (\ref{onepro})   
we find the following bound   
\beq \label{bounda}   
P_{\gamma \leftrightarrow a}(y) \le {(y \mu)^2\over 8 N}.   
\eeq   
As discussed in the appendix there are several assumptions that enter
the derivation of (\ref{totpro}). Still, we do not expect significant
deviation from the bound (\ref{bounda}) in more realistic cases.

We also consider cases where the mean free path of photons, $L_{mfp}$, 
is not very large. In such cases it takes an average of 
$N=3R^2/L_{mfp}^2$ random steps to escape from a region of radius $R$. 
Assuming that the magnetic field is constant over a region of order 
$L_{mfp}$, the conversion probability is given by Eq. (\ref{totpro}) 
with 
\beq   
y=N L_{mfp}={3R^2 \over L_{mfp}}, \qquad   
L_{decay}={2 L_{mfp} \over 3 P_1},    
\eeq   
Using (\ref{bounda}) we get the bound  
\beq \label{boundrr}    
P_{\gamma \leftrightarrow a}(y=R) \le {3 (R \mu)^2\over 8}\,.   
\eeq   
Note that the bound does not depend on the photon mean free path.   
   
In the following we will make use of the bounds (\ref{onepro}), 
(\ref{bounda}) and (\ref{boundrr}). 
Of course, they are useful only when the characteristic 
scale set by the magnetic field ($1/\mu$) is much larger than the 
relevant scales in the problem ($x$, $y$ and $R$). 
As it turns out this is indeed the situation in the cases we study.

\section{Photon--axion mixing inside Supernova Ia}   
 
We now turn to study axion--photon mixing inside the  
supernova and determine under which physical conditions there is an  
appreciable photon--axion oscillation probability.  SNe Ia are  
identified by the variations of their apparent luminosity with  
time. The intrinsic peak luminosity of these SNe varies within  
$20\,\%$, and based on the time variation even this variation  
can be corrected for. This is the reason that they are used as  
``standard candles" for various cosmological tests. The spectral  
analysis of SNe Ia shows the absence of hydrogen. Supernovae of this  
type are believed to result from the thermonuclear explosion of White  
Dwarfs (WDs) when their masses are very close to the Chandrasekhar  
limit \cite{sarkar,pbpal}.  
   
In order to estimate the axion--photon oscillation probability inside 
SNe Ia we consider a simple model \cite{pinto}. The supernova is  
assumed to be a sphere of uniform density with an initial radius  
$R_0 \sim 10^9\;{\rm cm}$ that expands with an outer velocity 
of $v=c/30=10^9\;{\rm cm \;s}^{-1}$. Photons are emitted uniformly 
throughout the volume of the supernova and the energy distribution 
follows Planck's law \cite{pinto,eastman}.  The peak luminosity occurs 
within 10 to 20 days after the explosion. The progenitor stars are 
magnetized objects. Some WDs are known to host magnetic fields of 
intensity ranging between $10^5$ and $10^9\,$G. The magnetic fields 
during the supernova event are not very well understood. It is 
possible that due to the turbulent motions prior to the explosion the 
WD goes through a stage which might dramatically change the intensity 
of its initial magnetic field. In particular, if the effect is to 
increase the magnetic field its value can be as large as $B_0\sim 
10^{11}\,$G \cite{bolometric}. After the explosion of the WD, the 
ejecta undergo a large expansion. In a homologous expansion all 
components of the field decrease like 
\be \label{magexp}   
{B \over B_0} = {(R_0)^2 \over (R)^2} \approx {(R_0)^2 \over (v t)^2}\,,  
\label{magfield}   
\ee   
where the approximation is valid for $vt \gg R_0$. This background 
magnetic field can be chaotic which is a likely result of the 
turbulent motion prior to the explosion. 
   
We first consider the production of axions in the SN envelope due to 
photon conversion in its magnetic field.  The photon mean free path at 
peak luminosity is not well known.  Ref. \cite{eastman}, for example, 
estimates that $L_{mfp}$ is in the range $10^{6} - 10^{14}\,{\rm cm}$. 
The SN radius at that time, which is of order $10^{15}\,{\rm cm}$, is 
larger than the mean free path.  The conversion probability at peak 
luminosity is therefore bounded by Eq. (\ref{boundrr}). Using 
(\ref{magexp}), we get 
\beq   
P_{\gamma \leftrightarrow a}    
\le {3 \over 8} \left({B_0 R_0^2 \over M v t}\right)^2\approx 3 
\times 10^{-8}\;   
\left({B_0 \over 10^{11}\;{\rm G}}\right)^2   
\left({10\;{\rm days} \over t}\right)^2   
\eeq   
where for the last step we used $M=4 \times 10^{11}\;{\rm GeV}$,
$R_0=10^9\;{\rm cm}$ and $v=c/30$.  We learn that for values of $t \ge
10\;$days and $B_0\le 10^{11}\,$G the conversion probability is
negligible. Our conclusions would not hold in the unlikely case that
initial magnetic fields in SNe Ia reach values significantly higher than
$10^{14}\,$G.

Let us now discuss other processes by which axions could in principle  
be produced inside the SN.  It is well known that axion production is  
potentially large in the first stage of  type II supernovae explosions  
\cite{Raffelt:wa}. One would then like to check whether a similar situation  
occurs in the type Ia SNe explosions. However, this is not the case 
since the temperature and the density of matter in the plasma of SNe 
Ia at its peak luminosity are relatively very small.  Let us elaborate 
on this. 
  
The potentially relevant processes are the ones that occur at the 
typical energy scale of a few eV.  The axion luminosity due to these 
processes scales as $n_B^{p} n_e^r T^s$, where $n_B$ is the baryon 
density, $n_e$ the electron density, $T$ the temperature, and
$p$, $r$, and $s$ are process dependent nonnegative numbers. We consider 
the following processes: axiorecombination --- i.e. the reaction, where 
axions are emitted when electrons and ions form a bound state; axion 
bremsstrahlung from the electrons and nuclei in the plasma; Compton 
scattering with axion instead of a photon in the final state; and the 
axion emission from electrons via the conversion of longitudinal 
plasmons. The luminosity due to the axiorecombination scales as $n_B 
n_e T^{3/2}$\cite{Dimopoulos:1986kc}, the luminosity due to the 
bremsstrahlung from electrons scales as $n_B n_e 
T^{5/2}$\cite{Raffelt:wa}. The axion bremsstrahlung emission from 
nuclei in the eV energy range is suppressed by additional powers of 
$m_e/m_N$. The luminosity due to the Compton scattering scales as $n_e 
T^6$\cite{Raffelt:wa}, and the luminosity due to the plasmon 
conversion scales as $n_e^2 T$\cite{Mikheev:1998bg}.  The density of 
plasma in SNe Ia at the time of peak luminosity is of order 
$10^{-13}\;{\rm g}\, {\rm cm}^{-3}$ and the temperature is about $2 
\times 10^4 \,K$ \cite{eastman}. To get a rough estimate of the axion 
production in the SNe Ia we compare these values with the typical 
density $\rho \sim 10^{13} \;{\rm g}\,{\rm cm}^{-3}$ and the 
temperature of several MeV inside the core of SNe II 
\cite{Raffelt:wa}. Multiplying by the volume of the SNe explosion  
remnant and taking $n_{e,B} R^3$ roughly constant we arrive at a  
suppression factor of about $10^{-40}-10^{-50}$. This expectation is  
borne out by a more detailed calculation, where it is found that the  
axion flux due to these processes in the SNe Ia remnant is suppressed  
by about 50 orders of magnitude compared to the photon flux. We  
conclude that axion production by these mechanisms can be safely neglected.  
   
\section{Photon--axion mixing in the host Galaxy}   
   
Next we consider photon axion conversion in the host galaxy.  The 
details of the background magnetic field of the galaxy are not well 
understood. We first assume that far away galaxies are similar to the 
better known nearby galaxies. The size of the galactic disk, which is 
also the coherence length of the smooth component of the magnetic 
field is of order $1\; {\rm kpc} \sim 10^{26}~{\rm eV}^{-1}$. For a 
typical galaxy, Ref. \cite{carlson} estimates a smooth toroidal 
component of the magnetic field of magnitude about $2\,\mu$G and a 
random component with a magnitude of about $5\,\mu$G and a coherence 
length of order $50\; {\rm pc}$.  The mean electron density in the 
host galaxy is taken to be $n_e = 0.03~{\rm cm}^{-3}$ 
\cite{carlson,harari}.  With these parameters we find that the    
mean free path of photons is much larger than $1\;{\rm kpc}$ and thus 
they basically do not interact after they leave the SN.  For the 
smooth toroidal component of magnetic field, using $B=2\,\mu$G, we 
have $\mu \sim 10^{-28}\;{\rm eV}$.   
Using (\ref{onepro}) with $x=L_{dom}=1\; {\rm kpc}$ 
we conclude 
\be   
P_{\gamma \leftrightarrow a}  \lsim 10^{-4}.   
\ee   
The random component of the magnetic field is somewhat larger in 
strength. However, due to smaller coherence length its contribution to 
the conversion probability is likely to be even smaller. For example, 
using (\ref{bounda}) with $L_{dom}=50\; {\rm pc}$, $y=1\; {\rm kpc}$ 
and $B=5\,\mu$G we find 
\be   
P_{\gamma \leftrightarrow a}  \lsim 10^{-5}.   
\ee  
   
Far away host galaxies may have considerably different magnetic field 
than what we find in nearby galaxies. One or two order of magnitude 
enhancement of the magnetic field could result in a significant 
conversion.  (In order for such conversion to be realized also the 
plasma frequency has to be small.)  Yet, there is no well motivated
reason to believe that this is the case. Thus it is likely that the 
conversion probability in the far away host galaxy is small. 
   
\section{Conclusions}   
 
We have studied axion production in type Ia supernovae and
photon--axion oscillations in their magnetic fields and in the
magnetic fields of their host galaxies. If such axion production is
significant it can disfavor the photon--axion oscillation explanation
of the observed dimming of light from far away type Ia supernovae
since then the dimming is less significant, and it may also become
energy dependent. We found that all these effects are very small. Our
result is robust in the sense that it does not rely on many unknown
parameters of the SNe and their host galaxies. For example, we did not
have to use exact values of the plasma frequency or the photon mean
free path. We had to use only the rough size of the star, its
temperature and its magnetic field. We therefore conclude that axion
production in the supernovae and their host galaxies has a negligible
effect on the CKT explanation of the supernovae dimming.
   
\acknowledgments    
 
We are grateful to Borut Bajc, Adam Burrows, Arnon Dar, Yossi Nir, Yael
Shadmi and Eli Waxman for helpful discussions. J.Z. thanks the
Technion theory group for hospitality while this work was in
progress. Y.G.~was supported in part by the Israel Science Foundation
under Grant No.~237/01-1 and by the United States--Israel Binational
Science Foundation (BSF) through Grant No.  2000133. J.Z. was
supported in part by the Ministry of Education, Science and Sport of
the Republic of Slovenia.
    
 
\appendix  
\section{Conversion in varying background}   
 
In this appendix we discuss photon to axion conversions in the varying  
magnetic field background. To simplify the discussion it is assumed  
that photons and axions traverse $N$ domains of equal length. In  
each domain the magnetic field $B$ is assumed to be homogeneous with a  
discrete change in $B$ from one domain to the other, i.e.  the  
situation is far from adiabatic conditions. The component of the magnetic  
field perpendicular to the direction of flight is assumed to have  
random orientation but is equal in size in each domain.  The domains  
can be (not necessarily) separated by an interval where the conversion  
is negligible.  
   
Consider an initial state $c_1(0) |\gamma_1\rangle  
+c_2(0)|\gamma_2\rangle +c_a(0)|a\rangle $. The two photon  
polarization states, $|\gamma_{1,2} \rangle$, correspond respectively   
to photons parallel and perpendicular to the magnetic field in the  
first domain. The initial photon and axion fluxes are then  
$I_\gamma(0)\sim |c_1(0)|^2 +|c_2(0)|^2$ and $I_a(0)\sim |c_a(0)|^2$  
respectively. In the $n$-th domain the magnetic field is tilted by  
an angle $\theta_n$ compared to the magnetic field in the first domain  
\beq  
|\gamma_\parallel^n\rangle=c_n|\gamma_1\rangle+s_n|\gamma_2\rangle\,,\qquad  
|\gamma_\perp^n\rangle=-s_n|\gamma_1\rangle+c_n|\gamma_2\rangle \,,  
\eeq   
or   
\beq  
c_1(y)=c_n c_\parallel^n(y)-s_n c_\perp^n(y)\,, \qquad  
c_2(y)=s_n c_\parallel^n(y)+c_n c_\perp^n(y) \,,  
\eeq   
where $c_n\equiv \cos\theta_n$ and $s_n\equiv \sin\theta_n$. The  
magnetic field mixes photons polarized parallel to the magnetic field  
with the axions according to the mass matrix given in  
(\ref{matrix}). We denote the transition elements for one domain  
$t_{\gamma\to \gamma}=\bra  
\gamma_\parallel^n(L_{dom})|\gamma_\parallel^n(0)\ket$ and similarly  
for $t_{\gamma\to a}$. The values of the transition elements are equal  
in each domain since the magnitude of the magnetic field has been assumed  
to be the same everywhere. The transition probability for photon to  
axion oscillation in one domain is $P_1=|t_{\gamma\to a}|^2$ with  
$P_1$ given in (\ref{probab}), while the survival probability of  
photon is $1-P_1=|t_{\gamma\to\gamma}|^2$. At the end of the $n$-th  
domain the photon and axion fluxes are  
\bea   
I_\gamma(n+1)&\sim& (1-P_1c_n^2 )|c_1(y_n)|^2+(1-P_1s_n^2)|c_2(y_n)|^2+  
P_1|c_a(y_n)|^2+\dots\\   
I_a(n+1)&\sim&  
P_1c_n^2 |c_1(y_n)|^2+P_1s_n^2|c_2(y_n)|^2+(1-P_1)|c_a(y_n)|^2+\dots   
\eea   
where dots represent terms that are proportional to $\cos\theta_n$,  
$\sin\theta_n$ or $\cos\theta_n\sin\theta_n$. We have defined  
$y_n=(n-1)L_{dom}$. Namely, the coefficients $c_{1,2,a}$ are taken at the  
beginning of the $n$-th domain.  
   
Next we assume that the transition probability in one domain is small 
i.e., $P_1\ll 1$, and that the direction of the magnetic field is 
random, i.e., $\theta_n$ is a random variable with 
$\theta_{n+1}-\theta_n\sim {\cal O}(1)$. In this limit $c_n^2$ and 
$s_n^2$ can be replaced by their average value $1/2$, while $c_n$, 
$s_n$ and $c_n s_n$ are averaged to zero. Using the fact that 
$I_\gamma(n)\sim |c_1(y_n)|^2+|c_2(y_n)|^2$ and $I_a\sim |c_a(y_n)|^2$ 
we arrive at 
\bea   
\pmatrix{I_\gamma(n+1)\cr I_a(n+1)}&=&\pmatrix{1-\frac{1}{2}P_1  
&P_1\cr \frac{1}{2}P_1&1-P_1}\pmatrix{I_\gamma(n)\cr I_a(n)} =\\ \cr  
&&\pmatrix{\frac{2}{3}+\frac{1}{3}\left[1-\frac{3  
}{2}P_1\right]^{n+1}& \frac{2}{3}-\frac{2}{3}\left[1-\frac{3  
}{2}P_1\right]^{n+1}\cr  
\frac{1}{3}-\frac{1}{3}\left[1-\frac{3 }{2}P_1\right]^{n+1}&  
\frac{1}{3}+\frac{2}{3}\left[1-\frac{3 }{2}P_1\right]^{n+1}}  
\pmatrix{I_\gamma(0)\cr I_a(0)} \,. \nonumber  
\eea   
We can then use the fact that the number of domains is large and  
replace $\left(1-3P_1/2\right)^{n+1}$ with the limiting  
expression $\exp\left[-(3 P_1 y)/(2L_{dom})\right]$ to  
arrive at the final expressions  
\bea   
I_\gamma(y)&=&I_\gamma(0)-P_{\gamma\to a}(I_\gamma(0)-2 I_a(0)) \,, \\   
I_a(y)&=&I_a(0)+P_{\gamma\to a}(I_\gamma(0)-2 I_a(0)) \,,  
\eea   
with   
\beq  
P_{\gamma\to a}=\frac{1}{3}\left(1-e^{-y/L_{decay}}\right)\,,  
\eeq  
and $L_{decay}=(2 L_{dom})/(3P_1)$. In cases where the   
oscillation length  
is much larger than the domain size the conversion probability $P_1$  
saturates the upper bound $P_1=(\mu L_{dom})^2/4$ and we  
arrive at the expression $L_{decay}=8/(3 \mu^2 L_{dom})$ given  
also in \cite{ckt1}.  
   
\def\pr#1,#2 #3 { {Phys.~Rev.}        ~{\bf #1},  #2 (19#3) }   
\def\prd#1,#2 #3{ {Phys.~Rev.}~{D \bf #1}, #2 (19#3) }   
\def\pprd#1,#2 #3{ {Phys.~Rev.}~{D \bf #1}, #2 (20#3) }   
\def\prl#1,#2 #3{ {Phys.~Rev.~Lett.}  ~{\bf #1},  #2 (19#3) }   
\def\pprl#1,#2 #3{ {Phys.~Rev.~Lett.}  ~{\bf #1},  #2 (20#3) }   
\def\plb#1,#2 #3{ {Phys.~Lett.}       ~{\bf B#1}, #2 (19#3) }   
\def\pplb#1,#2 #3{ { Phys.~Lett.}       ~{\bf B#1}, #2 (20#3) }   
\def\npb#1,#2 #3{ { Nucl.~Phys.}       ~{\bf B#1}, #2 (19#3) }   
\def\prp#1,#2 #3{ { Phys.~Rep.}       ~{\bf #1},  #2 (19#3) }   
\def\zpc#1,#2 #3{ { Z.~Phys.}          ~{\bf C#1}, #2 (19#3) }   
\def\epj#1,#2 #3{ { Eur.~Phys.~J.}     ~{\bf C#1}, #2 (19#3) }   
\def\mpl#1,#2 #3{ { Mod.~Phys.~Lett.}  ~{\bf A#1}, #2 (19#3) }   
\def\ijmp#1,#2 #3{{ Int.~J.~Mod.~Phys.}~{\bf A#1}, #2 (19#3) }   
\def\ptp#1,#2 #3{ { Prog.~Theor.~Phys.}~{\bf #1},  #2 (19#3) }   
   
   
\end{document}